\documentclass[trackchanges, twocolumn, twocolappendix]{aastex701}

\usepackage{enumitem}

\usepackage{color}

\shorttitle{Non-spherical Void Galaxies}
\shortauthors{Y.Song et al.}
\graphicspath{{./}{figures/}}
\usepackage{subfigure} 
\usepackage{amsmath}
\begin{document}

\title{Quantifying Environmental Effects on Galaxy Properties using Non-spherical Voids Identified from SDSS DR7}
\author[orcid=0009-0007-5626-9489]{Yingxiao Song}
\affiliation{Department of Astronomy, School of Physics and Astronomy, Shanghai Jiao Tong University, Shanghai 200240, People’s Republic of China}
\affiliation{Shanghai Key Laboratory for Particle Physics and Cosmology, and Key Laboratory for Particle Physics, Astrophysics and Cosmology, Ministry of Education, Shanghai Jiao Tong University, Shanghai 200240, People’s Republic of China}
\email[show]{yxsong@sjtu.edu.cn}  

\author[orcid=0000-0003-3997-4606]{Xiaohu Yang}
\affiliation{Department of Astronomy, School of Physics and Astronomy, Shanghai Jiao Tong University, Shanghai 200240, People’s Republic of China}
\affiliation{Shanghai Key Laboratory for Particle Physics and Cosmology, and Key Laboratory for Particle Physics, Astrophysics and Cosmology, Ministry of Education, Shanghai Jiao Tong University, Shanghai 200240, People’s Republic of China}
\affiliation{State Key Laboratory of Dark Matter Physics, Tsung-Dao Lee Institute, Shanghai Jiao Tong University, Shanghai 200240, People’s Republic of China}
\email[show]{xyang@sjtu.edu.cn}  

\author[orcid=0000-0003-0709-0101]{Yan Gong}
\affiliation{National Astronomical Observatories, Chinese Academy of Sciences,20A Datun Road, Beijing 100012, People’s Republic of China}
\affiliation{School of Astronomy and Space Sciences, University of Chinese Academy of Sciences(UCAS),\\Yuquan Road NO.19A Beijing 100049, People’s Republic of China}
\affiliation{Science Center for China Space Station Telescope, National Astronomical Observatories, Chinese Academy of Sciences,\\20A Datun Road, Beijing 100101, People’s Republic of China}
\email{gongyan@bao.ac.cn}

\author[orcid=0000-0003-0850-3641]{Haitao Miao}
\affiliation{National Astronomical Observatories, Chinese Academy of Sciences,20A Datun Road, Beijing 100012, People’s Republic of China}
\affiliation{Science Center for China Space Station Telescope, National Astronomical Observatories, Chinese Academy of Sciences,\\20A Datun Road, Beijing 100101, People’s Republic of China}
\affiliation{State Key Laboratory of Radio Astronomy and Technology, National Astronomical Observatories, Chinese Academy of Sciences, Beijing 100101, People's Republic of China}
\email{miaoht@bao.ac.cn}

\author[orcid=0000-0001-7283-1100]{Xingchen Zhou}
\affiliation{National Astronomical Observatories, Chinese Academy of Sciences,20A Datun Road, Beijing 100012, People’s Republic of China}
\affiliation{Science Center for China Space Station Telescope, National Astronomical Observatories, Chinese Academy of Sciences,\\20A Datun Road, Beijing 100101, People’s Republic of China}
\email{xczhou@nao.cas.cn}

\author[orcid=0000-0003-3196-7938]{Yizhou Gu}
\affiliation{Department of Astronomy, School of Physics and Astronomy, Shanghai Jiao Tong University, Shanghai 200240, People’s Republic of China}
\affiliation{Shanghai Key Laboratory for Particle Physics and Cosmology, and Key Laboratory for Particle Physics, Astrophysics and Cosmology, Ministry of Education, Shanghai Jiao Tong University, Shanghai 200240, People’s Republic of China}
\affiliation{State Key Laboratory of Dark Matter Physics, Tsung-Dao Lee Institute, Shanghai Jiao Tong University, Shanghai 200240, People’s Republic of China}
\email{guyizhou@sjtu.edu.cn}

\author[orcid=0000-0003-0939-9671]{Yingjie Peng}
\affiliation{Department of Astronomy, School of Physics, Peking University, Beijing 100871, People’s Republic of China}
\affiliation{Kavli Institute for Astronomy and Astrophysics (KIAA), Peking University, Beijing 100871, People’s Republic of China}
\email{yjpeng@pku.edu.cn}

\correspondingauthor{Yingxiao Song, Xiaohu Yang}
 
\begin{abstract}

Cosmic voids provide a distinct low-density region for studying the environmental effects of galaxy properties. Using the SDSS DR7 catalog, we identify non-spherical voids via Voronoi tessellation and the watershed algorithm, and classify void galaxies based on their local volume. We compare and find that void galaxies classified by this method are systematically less massive, fainter, bluer, and have higher specific star formation rate (sSFR) than non-void galaxies and all galaxy samples. We then divide void and non-void galaxies into stellar mass bins to focus on the environmental dependence of $g-r$ color and sSFR. By further classifying galaxies into blue/red and star-forming/quiescent populations, we calculate the ratio of blue to red and star-forming to quiescent for void and non-void galaxies separately. Comparing the ratio of the void value to the non-void value for both metrics presents an overall decreasing trend with stellar mass $M_*$ over the $9.4-10.4$ range in $\log[M_*/\mathrm{M}_\odot]$, indicating a stronger environmental effect in lower-mass systems. These results show that our classification of void galaxies in non-spherical voids based on local volume offers a robust approach for quantifying the influence of underdense environments on galaxy evolution.

\end{abstract}


\section {introduction} \label{sec:intro}

Cosmic voids, the vast underdense regions that dominate the volume of the Universe, are key components of the cosmic web along with filaments, sheets, and clusters \citep{1996Natur.380..603B}. Their low density and large size keep their evolution largely in the linear regime, making them powerful probes for constraining cosmological parameters \citep[e.g.][]{2022MNRAS.511.5492Z,2022A&A...667A.162C,contarini2023cosmological,2026A&ARv..34....1C,2023A&A...670A..47B,2023JCAP...08..010V,2023A&A...674A.185M,2023A&A...677A..78R,2024A&A...691A..39R,2024JCAP...10..079V,2024ApJ...970L..32W,2025ApJ...993..227V,2024MNRAS.534..128S,2024ApJ...976..244S,2026arXiv260222990L,2026arXiv260309278X,2026arXiv260329706Z}. 

Beyond cosmology, voids also provide an effective setting for galaxy formation 
and evolution studies. Compared with clusters and filaments, voids contain far fewer galaxies and much less matter. This makes them a uniquely different environment for studying how the large-scale structure influences galaxy properties. Studies from both simulations and observations have consistently found that void galaxies tend to be fainter, bluer, less massive, and have higher star formation rate (SFR) than galaxies in denser regions \citep[e.g.][]{2015ApJ...810..108M,2016ApJ...831..118M,2016MNRAS.458..394B,2024MNRAS.528.2822R,2024ApJ...962...58C,2026ApJ...998..325C}.

However, the robustness of these observed trends depends on how void galaxies are classified. Most previous studies of void galaxies have relied on spherical void catalogs \citep[e.g.][]{2023Natur.619..269D,2023ApJ...958...59V,2025ApJ...978....3Z,2026arXiv260409983P}. This is because the identification of non-spherical voids, which uses Voronoi tessellation and the watershed algorithm \citep[e.g.][]{watershed,zobov,vide,vast,2025JCAP...12..001G}, can incorporate tracer particles from relatively high-density regions into the void boundaries. If we directly apply the galaxy classification method suitable for spherical voids to non-spherical void samples — a galaxy is assigned to a void if its distance to the void center is less than the void radius — we will misclassify some galaxies that do not reside in low-density environments. In fact, the void samples used in recent cosmological studies are largely based on non-spherical voids \citep[e.g.][]{hamaus2016constraints,2022A&A...658A..20H,nadathur2019accurate,nadathur2020completed,2022MNRAS.516.4307W,2022MNRAS.513..186A,2024ApJ...969...89T,2025MNRAS.538..114S,2026ApJ..1007..169S}. 
Hence, we propose to classify galaxies in non-spherical voids based on the volume of the Voronoi cell assigned to each galaxy in the Voronoi tessellation. This cell volume reflects the local volume around the galaxy, so a larger volume indicates a lower density environment for that galaxy. 
We then compare the properties of void galaxies identified in our method with those of non-void galaxies to characterize the differences between the galaxy population in low-density environments and the remaining galaxy population.

Furthermore, while numerous earlier investigations have reported that void galaxies are bluer and exhibit elevated star formation rates \citep[e.g.][]{2004ApJ...617...50R,2005ApJ...624..571R,2007ApJ...658..898P,2012AJ....144...16K,2012MNRAS.426.3041H,2021ApJ...906...97F}, the observed contrasts in color and specific star formation rate (sSFR), defined as the SFR per unit stellar mass, between void and non-void galaxies may primarily reflect the typically lower stellar masses of void galaxies, rather than a direct influence of the environment. If the color and sSFR distributions of void and non-void galaxies are compared in bins of stellar mass, the residual differences can be largely attributed to environmental effects rather than to the intrinsic color-stellar mass and sSFR-stellar mass relations. To disentangle the roles of intrinsic and external factors, we evaluate how environment affects two galaxy properties, color and sSFR, by controlling for stellar mass, thereby minimizing the influence of intrinsic characteristics within the galaxy samples.

In this work, we first identify non-spherical void samples with Voronoi tessellation and the watershed algorithm, using the galaxy catalog from the seventh data release (DR7) of the Sloan Digital Sky Survey \citep[SDSS,][]{2009ApJS..182..543A}. Since these voids can take arbitrary shapes and are not constrained to spherical regions, we flag the void galaxies based on the local volume of each galaxy in our catalog. Then we compare the number distributions of the four properties from the void galaxies, non-void galaxies and all galaxy samples, including stellar mass, luminosity, color, and sSFR. Next, we apply the $V_{\rm max}$ method to our galaxy samples and obtain the number density distributions of the $g-r$ color and the sSFR. We divide the samples into different stellar mass bins and, within each bin, compute the blue cloud to red sequence ratio from the $g-r$ color distribution, and the star-forming to quiescent ratio from the sSFR distribution.

The paper is organized as follows: In Section \ref{sec:data}, we introduce the galaxy catalog of SDSS DR7, the void finder we use, and the method of classification for void galaxies; In Section \ref{sec:vgp}, we present and discuss the property distributions of void galaxies, non-void galaxies, and all galaxy samples; Section \ref{sec:env} quantifies environmental effects through the blue cloud to red sequence ratio and the star-forming to quiescent ratio. The summary and conclusion are given in Section \ref{sec:conclusion}.

\section{Data Analysis} \label{sec:data}

\subsection{Galaxy Catalog} \label{sec:galcat}
\setcounter{footnote}{0}

We use the galaxy catalog from the SDSS DR7 for our analysis \citep{2009ApJS..182..543A}, which provides a flux-limited spectroscopic sample complete to an $r$-band apparent magnitude $m_r < 17.77$ \citep{2002AJ....124.1810S}. 
We restrict our analysis to the major contiguous region within the Northern Galactic Cap (NGC) to ensure a complete void identification. We adopt the redshift range $z = 0-0.114$, which is consistent with previous void studies from SDSS DR7 \citep[e.g.][]{2023ApJS..265....7D}. This region covers a sky area of about 7,300 deg$^2$ and contains approximately 380,000 galaxies.

The version of our galaxy catalog comes from the NASA-Sloan Atlas\footnote{\url{https://www.sdss4.org/dr17/manga/manga-target-selection/nsa/}} (NSA, version 1.0.1), since it directly provides stellar masses and $K$-corrected absolute magnitudes in the five photometric bands, i.e. $u$, $g$, $r$, $i$, $z$ \citep{2011AJ....142...31B}. 
The SFR and sSFR values we use are directly taken from the MPA-JHU value-added catalog\footnote{\url{https://wwwmpa.mpa-garching.mpg.de/SDSS/DR7/}} \citep{2004MNRAS.351.1151B}. Note that the galaxy catalog we use in our analysis was obtained by matching the NSA catalog with the MPA-JHU catalog using the RA and DEC coordinates with a precision of 1 arcsec. We resolved duplicate matches by choosing the one with the smallest angular distance, and about 9\% of the galaxies remained unmatched. We restricted our sSFR analysis to the matched galaxies in our catalog with valid sSFR measurements, and we have also verified that the normalized sSFR number distribution of this subset is consistent with that of the full MPA-JHU catalog.

In this work, we assume a flat $\Lambda$CDM cosmology based on $\it Planck$ 2018 results \citep{2020A&A...641A...6P}, with $\Omega_{\rm m}= 0.315$ and $H_0 = 100\,h\,\mathrm{km\,s^{-1}\,Mpc^{-1}}$, to compute comoving coordinates and the relevant properties of galaxies and voids. Note that we calculate the mean galaxy separation (MGS) defined as ${\rm MGS}(z) = \bar{n}_{\rm g}(z)^{-1/3}$, where $\bar{n}_{\rm g}(z)$ is the number density of galaxies in each redshift bin. And we show the MGS($z$) values derived from our galaxy catalog in Figure~\ref{fig:mgs}.

\subsection{Void Catalog} \label{sec:voidcat}

We identify non-spherical voids from the SDSS DR7 galaxy sample using the 
Void IDentification and Examination toolkit\footnote{\url{https://bitbucket.org/cosmicvoids/vide\_public/src/master/}} \citep[\texttt{VIDE},][]{vide}. \texttt{VIDE} applies Voronoi tessellation and the watershed algorithm \citep{watershed} to find low-density regions, based on the ZOnes Bordering On Voidness method \citep[\texttt{ZOBOV},][]{zobov}. The identified voids do not assume any particular shape, which matches the non-spherical void definition we focus on in this work. 
Although \texttt{VIDE} can further merge adjacent underdense zones which are identified by the watershed algorithm based on the density at their boundaries, here we adopt the unmerged zones as our void sample to avoid the void-in-void issue, and we only retain voids whose central density, measured within one quarter of their size, is below a threshold close to zero (i.e. $10^{-9}$).

\begin{figure}
\centering
\includegraphics[width=\columnwidth]{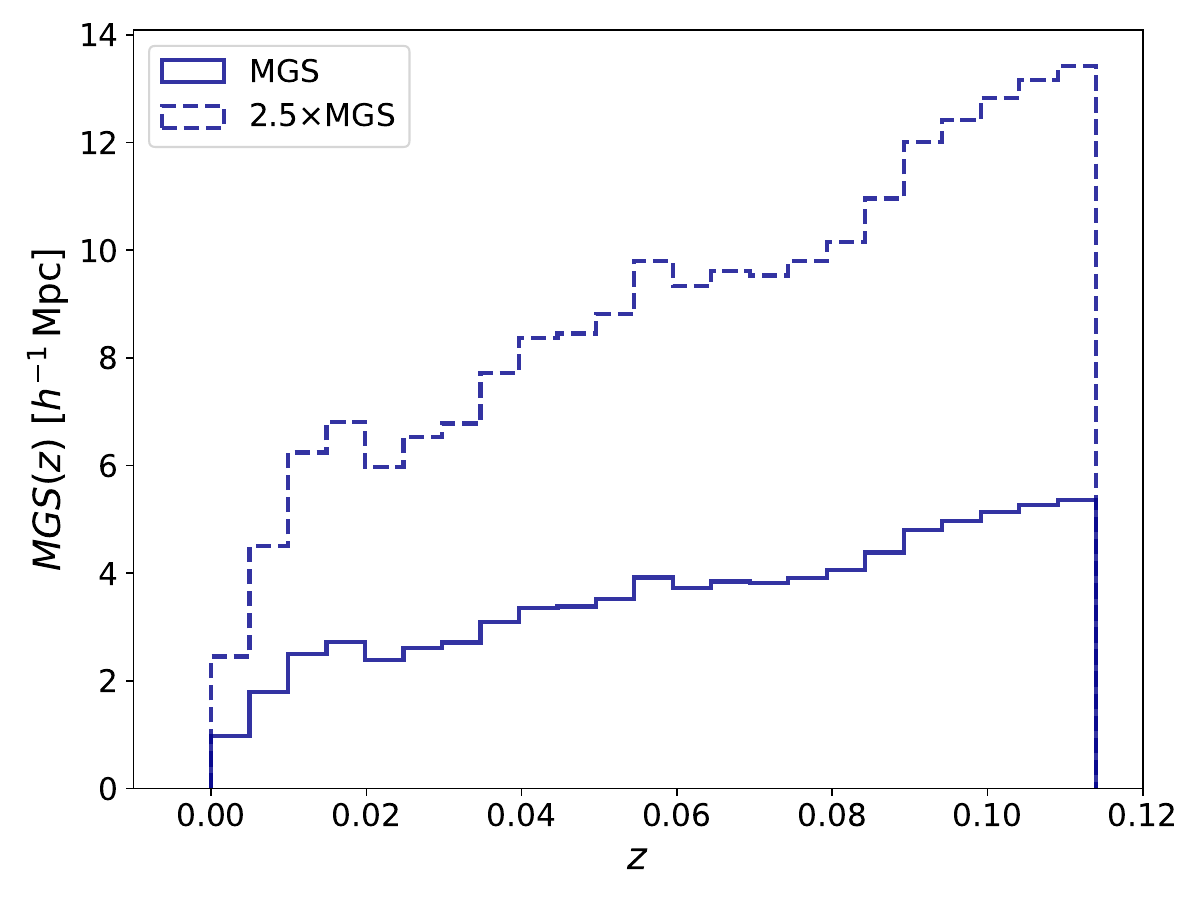}
\caption{The mean galaxy separation MGS($z$) (solid) and the corresponding 2.5$\times$MGS($z$) threshold used for void selection (dashed), derived from our SDSS DR7 galaxy catalog at $z= 0-0.114$. The redshift bins have a width of $\sim0.005$.}
\label{fig:mgs}
\end{figure}

\begin{figure*}
\subfigure{
\includegraphics[width=\columnwidth]{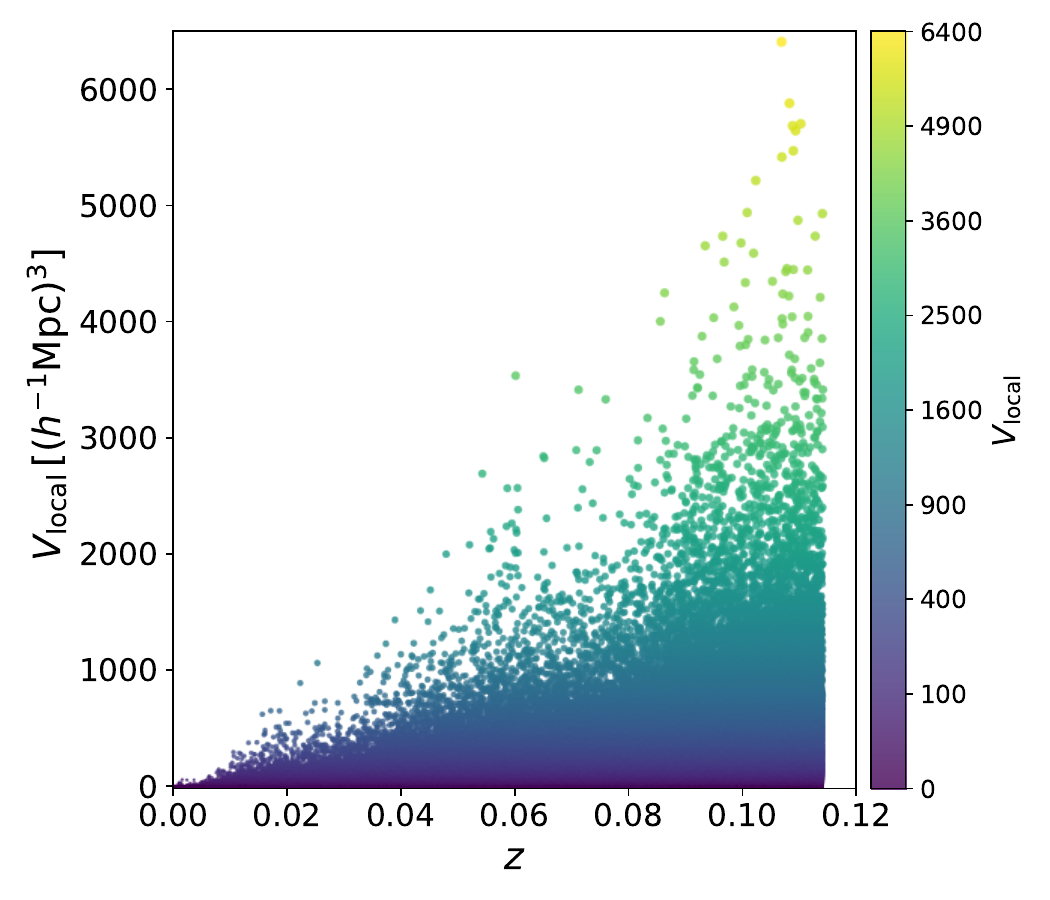}}
\hspace{2mm}
\subfigure{
\includegraphics[width=\columnwidth]{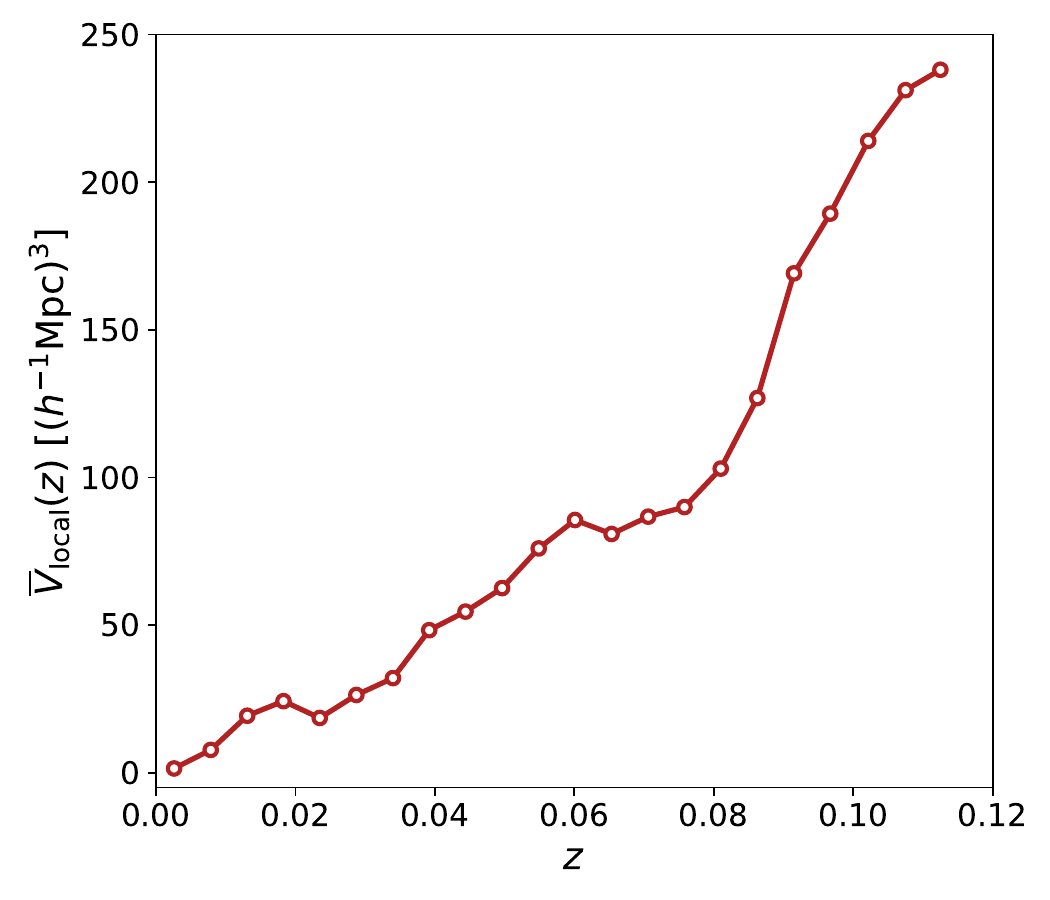}}
\caption{Left panel: The local volume distribution of our catalog from the SDSS DR7 in $0<z\leq0.114$. The color (from purple to yellow) and the size of the dots (from small to large) both scale with $V_{\rm local}$. 
Right panel: The mean local volume as a function of redshift for our SDSS DR7 sample at $ z\leq0.114$. The $\bar{V}_{\rm local}(z)$ is computed in redshift bins with a width of $\sim0.005$.}\label{fig:Vlocal}
\end{figure*}

To classify galaxies in non-spherical voids, it is necessary to understand how density is defined during the void identification process. Voronoi tessellation assigns a cell to each galaxy in the sample, such that every point within a given cell is closer to that galaxy than to any other. The volume of this cell, $V_{\rm cell}$, reflects the local volume $V_{\rm local}=V_{\rm cell}$ around that galaxy. The local number density is then estimated as $\rho_{\rm local} = 1/V_{\rm local}$, so that a larger $V_{\rm local}$ indicates a lower density environment. Note that we do not apply any cut on absolute magnitudes to our galaxy sample, meaning that we identify voids without considering a volume-limited sample. This is because we need to estimate the local volume for every galaxy in our catalog.  

For a given void, the total volume is the sum of the Voronoi cell volumes of the galaxies it contains, $V = \sum_{i=1}^N V_{\rm cell}^i$, where $N$ is the number of cells assigned to that void. The effective radius is defined as $R_{\rm v} = (3V/4\pi)^{1/3}$, and the volume-weighted center is $\mathbf{X}_{\rm v} = \sum_{i=1}^N \mathbf{x}_i V_{\rm cell}^i / V$, where $\mathbf{x}_i$ is the comoving position of the galaxy in the $i$-th cell.

To further ensure the completeness of our void samples, we only consider voids with an effective radius larger than 2.5 times the MGS($z$) at the corresponding redshift, where ${\rm MGS}(z)$ is computed at the bin centers of redshift bins with a width of $\sim0.005$ over the redshift range $0<z<0.114$. This threshold is consistent with the selection adopted in previous cosmological void studies \citep[e.g.][]{2019MNRAS.488.5075R,2019MNRAS.488.3526C,contarini2021cosmic,contarini2023cosmological,2024A&A...682A..20C,2025MNRAS.540.2853S}. The 2.5$\times$MGS($z$) values are also shown in Figure~\ref{fig:mgs}. After this selection, the void sample used in our analysis contains 573 voids, with the maximum effective radius reaching $\sim 33 \, h^{-1}{\rm Mpc}$.

\subsection{Identification of Void Galaxies} \label{sec:vg}

A common approach to identify void galaxies is to select galaxies within a certain fraction of the void radius from the void center. For non-spherical voids, the effective radius and volume-weighted center do not always capture the true extent and location of the underdense region. Even voids with relatively low ellipticity do not closely approximate a sphere, and smaller voids tend to exhibit higher ellipticities \citep[e.g.][]{2023ApJ...955..131W,2024MNRAS.532.1049S}. Given that the mean effective radius of our void sample is $17.1 \, h^{-1}{\rm Mpc}$, a simple distance-to-center cut will misclassify real void galaxies or include galaxies in overdense boundary regions.

We therefore require that a void galaxy must belong to one of the voids in our catalog: a galaxy is considered only if the Voronoi cell that hosts it is contained within one of the voids identified by \texttt{VIDE} and this void must satisfy the minimum radius cut based on MGS($z$) described in Section~\ref{sec:voidcat}.

However, galaxy membership in a void alone does not confirm that a galaxy resides in an underdense region, since many galaxies are located near the overdense boundaries of a non-spherical void. To ensure this, we rely on the local volume $V_{\rm local}$ to further select galaxies in genuinely low-density environments. We first compute the mean local volume $\bar{V}_{\rm local}(z)$ as a function of redshift from the full galaxy sample. A galaxy is considered to be located in a low-density environment if its local volume $V_{\rm local}$ exceeds $f \times \bar{V}_{\rm local}(z)$, where $\bar{V}_{\rm local}(z)$ is the mean local volume at that redshift and $f$ is a multiplicative factor. In Figure~\ref{fig:Vlocal}, we show the distribution of the local volume for all galaxies in our sample, along with the mean local volume as a function of redshift. We find that the largest $V_{\rm local}$ among all galaxies in our SDSS DR7 sample at $z\leq0.114$ is approximately 6,400 $h^{-3}{\rm Mpc}^3$.

Overall, we classify a galaxy as a void galaxy if it satisfies both conditions: its local volume meets $V_{\rm local}> f \, \bar{V}_{\rm local}(z)$, and it is associated with a void in our catalog. All other galaxies are classified as non-void galaxies in our analysis.

We note that, redshift-space distortions (RSD) might affect the process of non-spherical void identification and the estimated local volume $V_{\rm local}$ for galaxy samples in survey data. However, since we exclude small voids by keeping only those with an effective radius larger than $2.5 \times {\rm MGS}(z)$ in our void catalog, and our analysis focuses mainly on relatively isolated galaxies residing in these large voids, thus, our results are not expected to be significantly impacted by RSD effects. 

\section{Void Galaxy Properties}\label{sec:vgp}

\begin{deluxetable*}{cccccc}
\renewcommand{\arraystretch}{1}
\tablenum{1}
\tablecaption{The galaxy number and the median values of the $M_*$ (in log$[M_*/{\rm M}_\odot]$), $M_r$, $g-r$, and sSFR (in log$[{\rm yr}^{-1}]$) distributions for all galaxy sample and the void galaxy samples selected with $f = 1$, $2$, $3$ at $z \leq 0.114$ from our SDSS DR7 catalog. The errors indicate the $1\sigma$ uncertainties from jackknife method.} \label{tab:fmgs}
\setlength{\tabcolsep}{12pt} 
\tablehead{
\colhead{} &\colhead{Number}&
\colhead{$M_*$} & \colhead{$M_r$}&
\colhead{$g-r$}&
\colhead{sSFR}\\
&&\colhead{(Median)}&\colhead{(Median)}&\colhead{(Median)}&\colhead{(Median)}
}
\startdata
All Galaxies&378,209&$10.031 \pm 0.006$&$-20.030 \pm 0.013$&$0.635 \pm 0.002$&$-10.746 \pm 0.031$\\
\hline
Void Galaxies\\
\hline
$f$ = 1&51,624&$9.876 \pm 0.008$&$-19.765 \pm 0.019$&$0.563 \pm 0.002$&$-10.276 \pm 0.010$\\
$f$ = 2&26,496&$9.838 \pm 0.008$&$-19.700 \pm 0.021$&$0.549 \pm 0.003$&$-10.221 \pm 0.010$\\
$f$ = 3&15,435&$9.802 \pm 0.013$&$-19.651 \pm 0.019$&$0.539 \pm 0.003$&$-10.179 \pm 0.012$\\
\enddata
\end{deluxetable*}

To characterize the galaxy population in low-density environments, we focus on four properties: stellar mass, luminosity, color, and sSFR. Stellar mass $M_*$ drives the evolution of a galaxy, as it correlates with star formation efficiency, metal enrichment, and merger history. The luminosity can be described by the absolute magnitude, and we choose the $r$-band absolute magnitude $M_r$ as a representative measure of the optical luminosity. The color provides a measure of the relative contribution of young and old stellar populations, and we use $g-r$ results as an example for our analysis. The SFR quantifies how actively a galaxy is forming stars. We mainly consider the sSFR, which is the SFR per unit stellar mass and removes the dependence on stellar mass, to measure the star formation activity relative to the existing stellar content. In general, these four quantities allow us to examine both the stellar population and the star formation activity of galaxies in different density environments.

We measure the normalized number distributions of these four properties for the void galaxy samples constructed with the method described in Section~\ref{sec:vg}, where the normalisation is by the total number of galaxies in each sample. To assess how the choice of the local volume threshold $f$ affects the sample, we construct three void galaxy catalogs using $f = 1$, $2$, $3$. In Table~\ref{tab:fmgs}, we list the galaxy number and the median values of the $M_*$, $M_r$, $g-r$, and sSFR distributions for the three void galaxy samples and the full galaxy sample. We can find that the number of void galaxies decreases as $f$ increases, while the median offsets between the void and all galaxy samples become larger. This trend indicates that larger $f$ selects galaxies with larger local volumes, corresponding to more isolated galaxies. 
Here we adopt the jackknife method by dividing the full galaxy sample into 100 sub-samples to estimate the 1$\sigma$ uncertainties, and we find the median offsets between the all galaxies and void galaxy samples with $f=1,2,3$ are highly significant, exceeding $5\sigma$ for all four properties.
The normalized number distributions of these four properties from void galaxy samples with $f=1,2,3$ are shown in the Appendix \ref{sec:appendix1}.
We also test larger $f$ values and find that the offsets in median values of four galaxy property distributions relative to the full sample increase, but the rate of increase decreases with $f$.

\begin{figure*}
    \centering
    \subfigure{\includegraphics[width=\columnwidth]{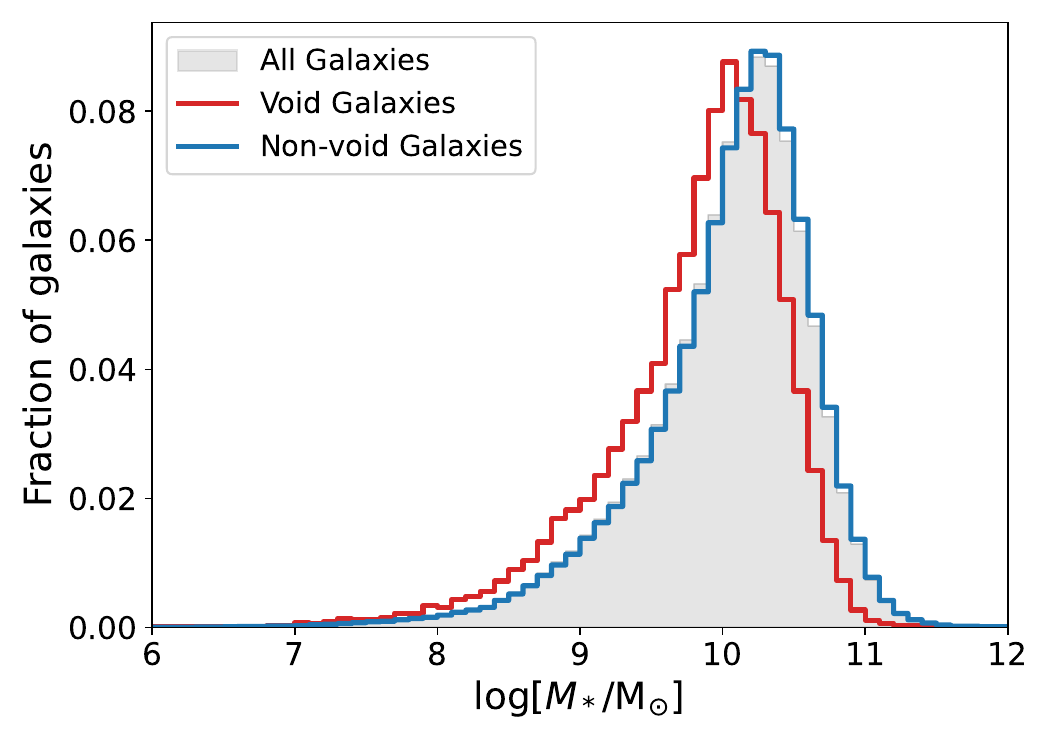}}
    \hspace{2mm}
    \subfigure{\includegraphics[width=\columnwidth]{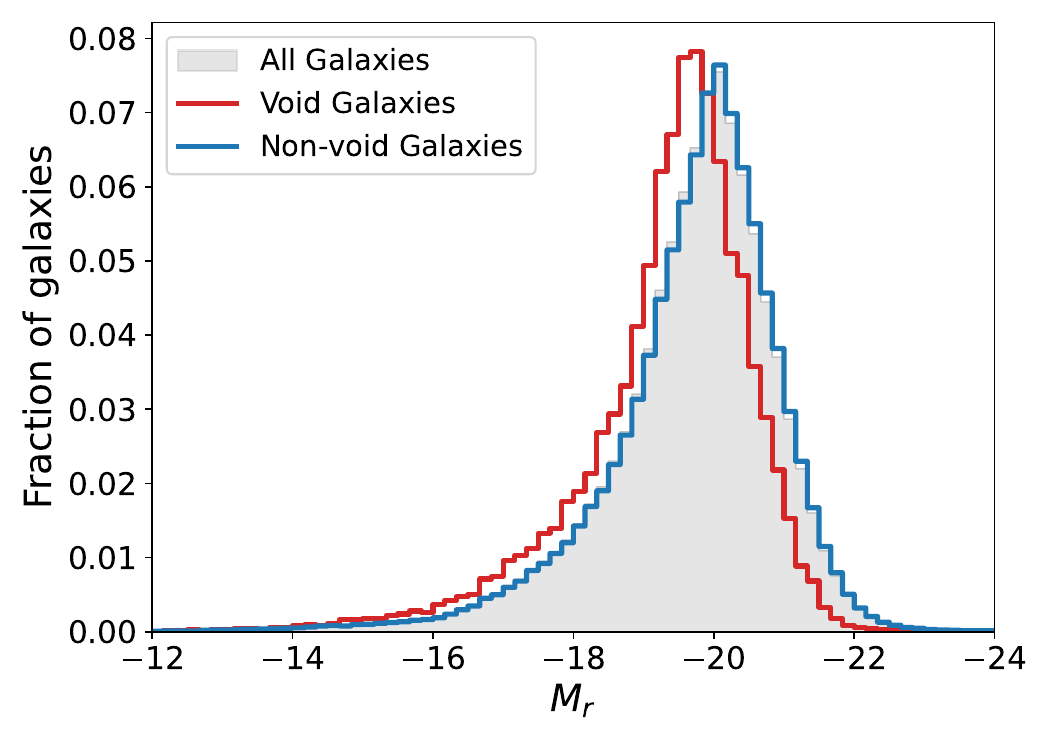}}
    \subfigure{\includegraphics[width=\columnwidth]{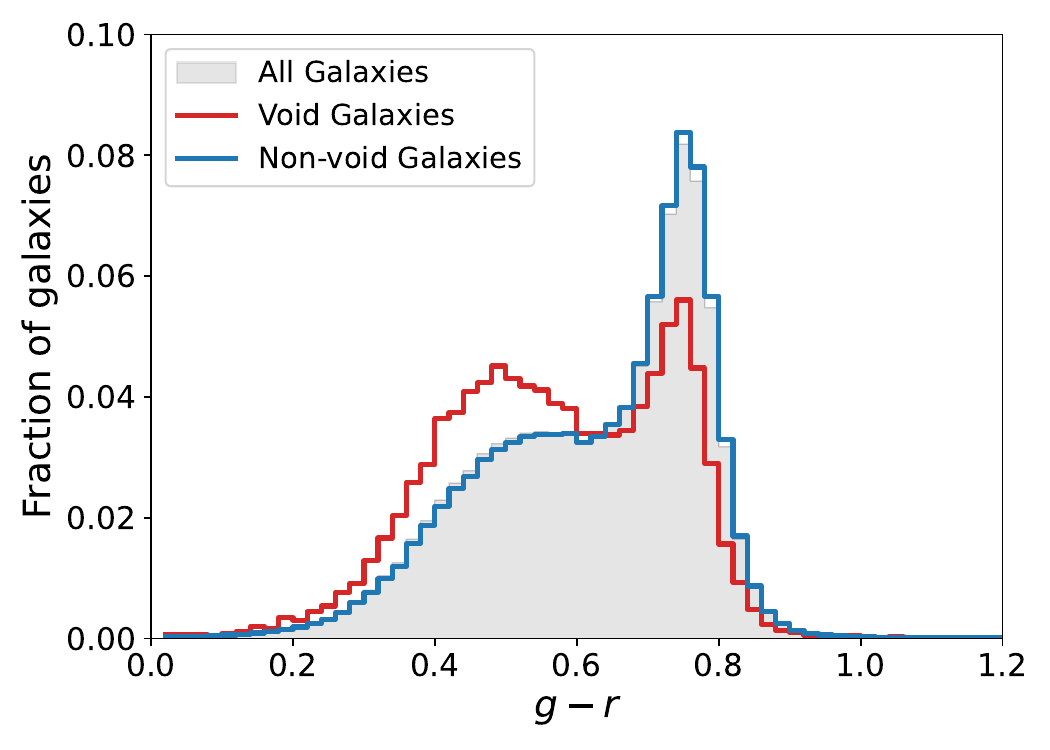}}
    \hspace{2mm}
    \subfigure{\includegraphics[width=\columnwidth]{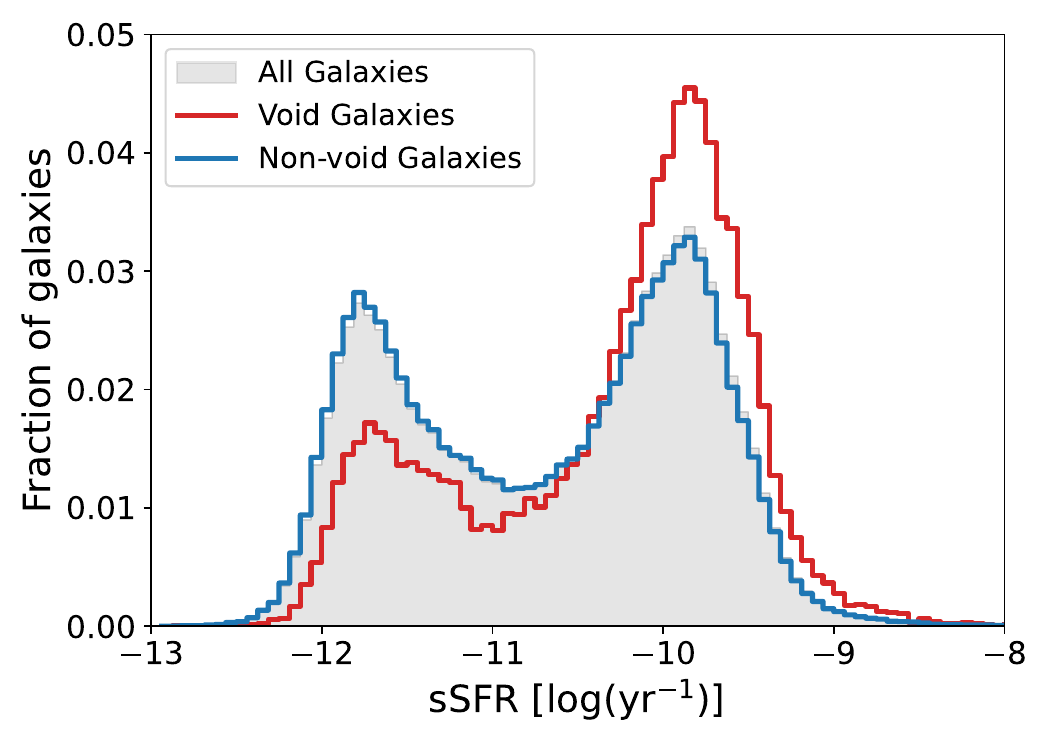}}
    \caption{The normalized number distributions of $M_*$, $M_r$, $g-r$, and sSFR for the void galaxy sample classified from our non-spherical void catalog with $f=2$ (red), the non-void sample (blue), and the all galaxy sample (gray shaded). Each distribution is normalized by the total number of galaxies in the corresponding sample.}\label{fig:vnd}
\end{figure*}

We find that, considering the number of void galaxies, the void galaxy sample with $f = 2$ provides a reliable choice. The void galaxy sample with $f = 1$ includes approximately 14\% of the full galaxy sample, which is likely to contain galaxies that are not sufficiently isolated, while the void galaxy sample with $f = 3$ includes only about 4\% of the full sample, and the number of void galaxies would be too small to provide reliable statistical results. 
Moreover, the choice of $f$ is defined relative to the mean local density $\bar{\rho}_{\rm local}(z)$, since $\bar{\rho}_{\rm local}(z)$ at different redshifts reflects the mean number density of galaxies $\bar{n}(z)$. Adopting $f=2$ corresponds to selecting galaxies with local densities $\rho_{\rm local} < 0.5\bar{\rho}_{\rm local}(z).$ We therefore adopt $f = 2$ as our fiducial threshold for the following analysis.
In Figure~\ref{fig:vnd}, we show the distributions of $M_*$, $M_r$, $g-r$, and sSFR for the void galaxy sample with $f=2$, the corresponding non-void sample, and the full galaxy sample. We can see clear offsets between the void galaxy sample and the other two samples in all four panels. We also classify void galaxies using a spherical void catalog constructed from the SDSS DR7 data (see the Appendix \ref{sec:appendix2} for details), and the relevant results show that the relative bias behaviors of the void galaxies and the other two samples are qualitatively similar to those we obtain from our non-spherical void analysis with $f=2$.

These findings indicate that the void galaxies classified by our method from non-spherical voids are on average less massive, fainter, bluer, and have higher sSFR than the non-void and full galaxy samples. 
These trends are consistent with previous observational results for galaxies in low-density environments \citep[e.g.][]{2025ApJ...978....3Z}. Therefore, these results demonstrate that our method of classifying void galaxies in non-spherical voids based on local volume provides a robust way to separate the low-density galaxy population from the general galaxy population, and also offers a reliable void galaxy sample for studying the environmental dependence of galaxy properties.

\section{Environmental Effects on Galaxy Properties}\label{sec:env}

In this section, we aim to investigate environmental impacts by dividing galaxies into red and blue, as well as star-forming and quiescent, sub-populations.

\subsection{Number Density Distributions of Galaxy Properties}\label{sec:vmax}

To accurately divide galaxies into distinct sub-populations, we begin by directly determining their color distributions and specific star formation rate distributions.
As emphasized in a recent study \citep{Wang2026}, in a flux-limited galaxy sample the observed galaxy number distribution can be significantly affected by survey selection effects; therefore, it is more appropriate to work with the galaxy number density.  
To account for the selection bias introduced by the flux limit of the SDSS DR7 sample, we adopt the $V_{\rm max}$ method to derive the number density distributions of two key galaxy properties ($g-r$ color and sSFR) for the void, non-void, and full galaxy samples. For all galaxy samples in our catalog, we compute the maximum observable volume $V_{\rm max}$ and weight each galaxy by $1/V_{\rm max}$ to construct the number density distribution of galaxy properties. The calculation uses the $r$-band apparent magnitude limit $m_r < 17.77$, the survey area of about 7,300 deg$^2$ and the redshift range $z = 0-0.114$, all of which are from the SDSS survey region we focus on.

\begin{deluxetable}{lccc}
\tablenum{2}
\tablecaption{Coefficients for the mean $r$-band $K$-correction in four color bins at $z=0-0.2$.}\label{tab:kmean}
\setlength{\tabcolsep}{15pt}  
\tablewidth{\textwidth} 
\tablehead{
& \colhead{$a_{\mu}$} &\colhead{$b_{\mu}$ } &\colhead{$c_{\mu}$}
}
\startdata
$\mu$ = 1 (blue)&$-$4.72 &1.48 &$-$0.22\\
$\mu$ = 2 &$-$1.52 &0.98&$-$0.19\\
$\mu$ = 3 &0.40 &0.78 &$-$0.19\\
$\mu$ = 4 (red)&1.10& 0.78 &$-$0.19\\
\enddata
\end{deluxetable}

\begin{figure*}
    \centering
    \includegraphics[width=\textwidth]{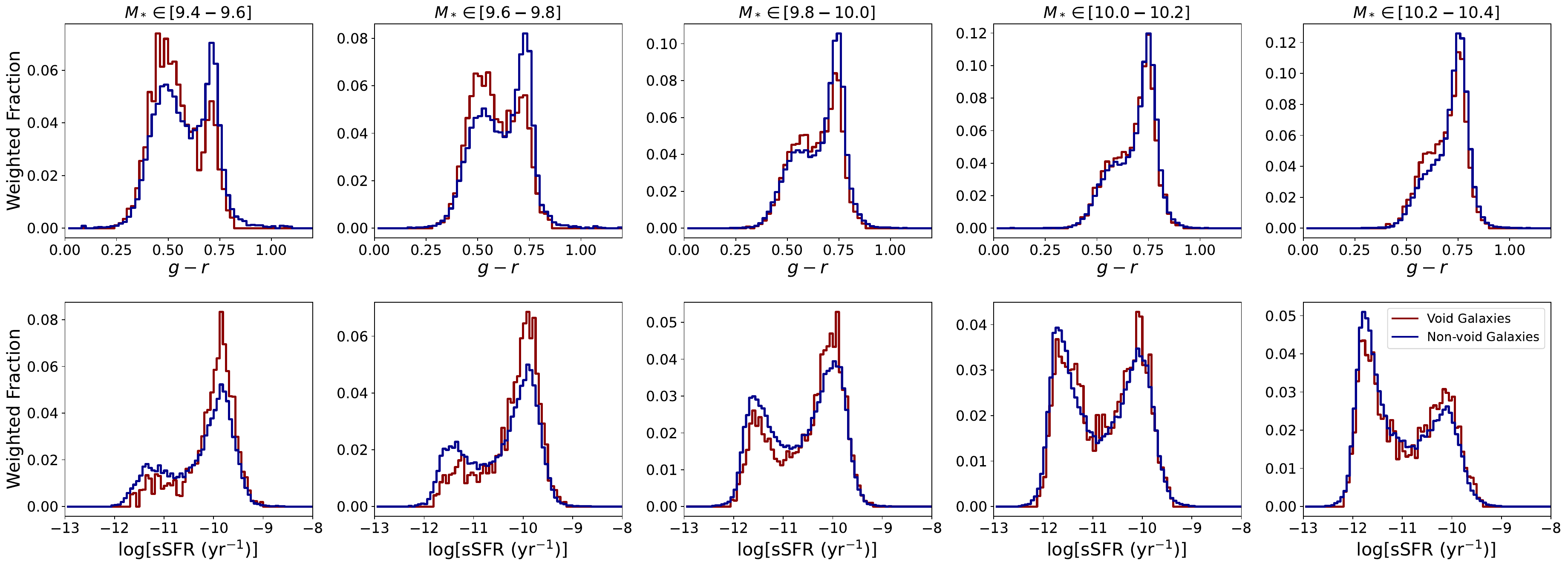}
    \caption{The normalized number density distributions of $g-r$ color (top) and sSFR (bottom) for void galaxies (dark red) and non-void galaxies (dark blue) across different stellar mass bins. These distributions are normalized to the total weight within each mass bin.}
    \label{fig:ndd}
\end{figure*}

To calculate $V_{\rm max}$, we first classify galaxies into four color bins according to their color $r - z$ within $z=0-0.114$, denoted by $\mu \in \{1,2,3,4\}$. For each color class $\mu$, the mean $K$-correction in the $r$-band as a function of redshift is approximated by
\begin{equation}
\bar{K}_{r,\mu}(z) = a_{\mu}z^2 + b_{\mu}z + c_{\mu}.
\end{equation}
The coefficients $a_\mu$, $b_\mu$, and $c_\mu$ are obtained by the results of \citet{2024ApJ...971..119W} over the redshift range $z=0-0.2$, which covers our sample redshift range $z=0-0.114$. The results of these coefficients for each color bin are listed in Table~\ref{tab:kmean}.

\begin{figure*}
    \centering
    \includegraphics[width=\textwidth]{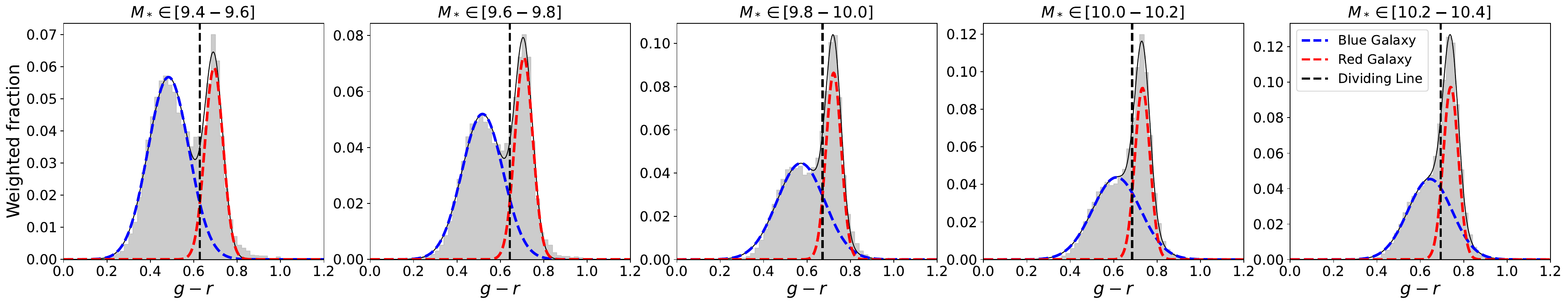}
    \caption{Classification of galaxies into blue cloud and red sequence populations across five stellar mass bins based on the normalized number density distribution of $g-r$ color. The gray shaded histogram represents the distribution of all galaxies from our SDSS DR7 catalog. The blue and red dashed lines show the bimodal Gaussian fittings for the blue cloud and red sequence populations, respectively. The black solid line indicates the sum of the two components, representing the total fit. The black vertical dashed line marks the dividing line that separates the blue and red populations.}
    \label{fig:grdl}
\end{figure*}

\begin{figure}
\centering
\includegraphics[width=\columnwidth]{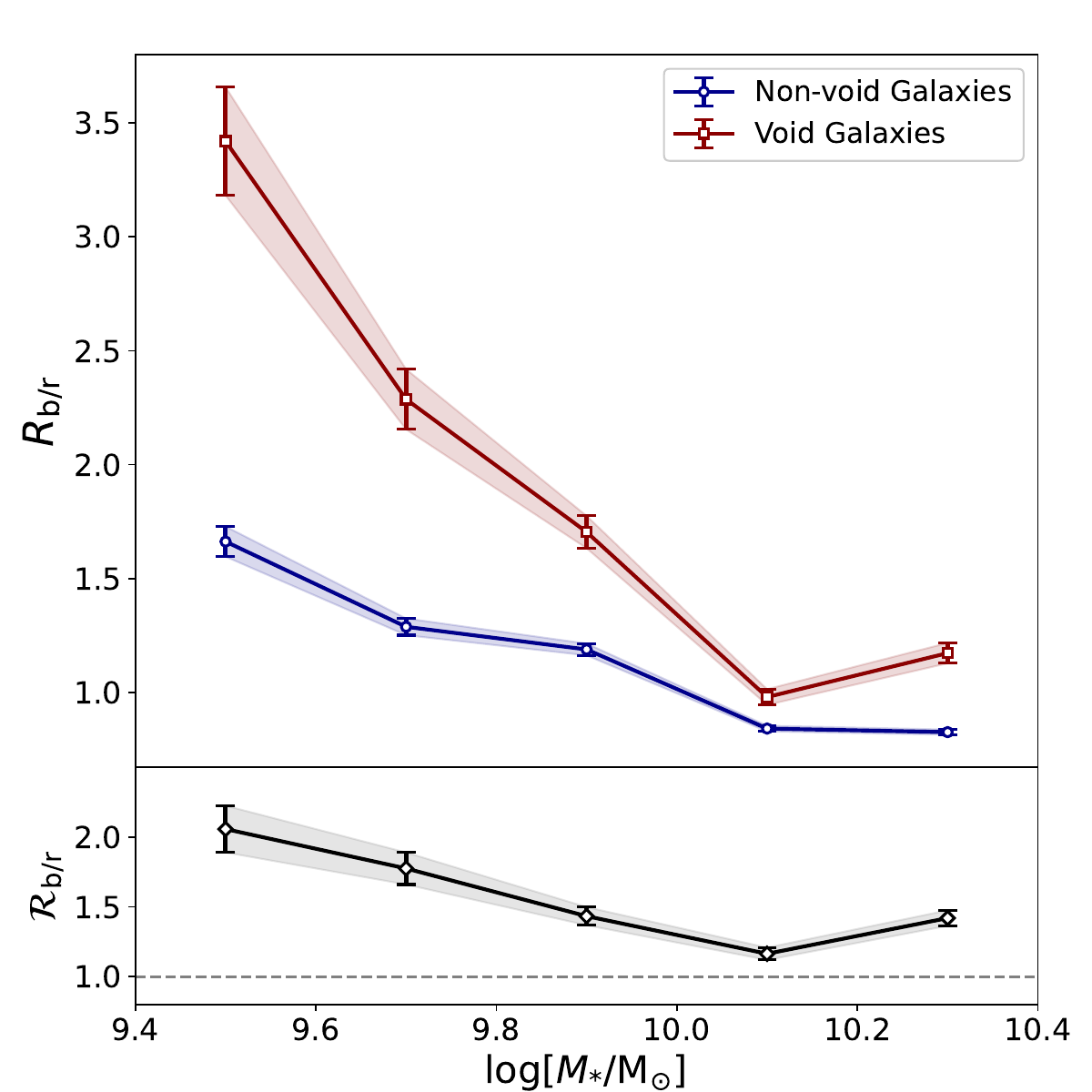}
\caption{The ratio of blue to red galaxies $R_{\rm b/r}$ as a function of stellar mass for void galaxies (dark red) and non-void galaxies (dark blue). The bottom panel shows the ratio of these two values, $\mathcal{R}_{\rm b/r}$. The error bars and shaded regions indicate the $1\sigma$ uncertainties.}
\label{fig:rbr}
\end{figure}

For a galaxy observed at $z_{\rm obs}$, its $K$-correction at $z_{\rm max}$ is estimated from the measured $K$-correction $K_r(z_{\rm obs})$ by applying the color-dependent mean correction
\begin{equation}
K_r(z_{\rm max}) = K_r(z_{\rm obs}) + \bar{K}_{r,\mu}(z_{\rm max}) - \bar{K}_{r,\mu}(z_{\rm obs}).
\end{equation}
For each galaxy with a known absolute magnitude $M_r$, the maximum redshift $z_{\rm max}$ is determined by solving
\begin{equation}
m_r^{\rm lim} - M_r = 5 \log_{10} \left[ \frac{(1+z_{\rm max}) D_{\rm cov}(z_{\rm max})}{10~\mathrm{pc}} \right] + K_r(z_{\rm max}),
\end{equation}
where $m_r^{\rm lim}=17.77$ and $D_{\rm cov}(z_{\rm max})$ is the comoving distance at $z_{\rm max}$. Since the lower redshift limit of our sample is $z=0$, the upper redshift limit for the volume calculation is $z_{\rm upper} = \min(z_{\rm max}, 0.114)$.
The maximum observable volume $V_{\rm max}$ is then computed by
\begin{equation}
V_{\rm max} = \frac{A_{\rm survey}}{A_{\rm sky}} \cdot \frac{4\pi}{3} D_{\rm cov}^3(z_{\rm upper}),
\end{equation}
where $A_{\rm survey}\simeq 7,300 \ {\rm deg}^2$ is the survey area of the SDSS data we focus on and $A_{\rm sky}=41,253\ \mathrm{deg}^2$ is the total area of the sky.

Once we obtained $V_{\rm max}$ for all our galaxies, we divide them into stellar mass bins with a width of 0.2. 
Then we compute the number density distributions across the stellar mass range from 9.0 to 11.0 in $\log[M_*/\mathrm{M}_\odot]$. To ensure statistical robustness, we restrict our analysis to mass bins containing more than 2,000 void galaxies, which results in the adopted range of $9.4-10.4$. 
Within each mass bin, we apply the $V_{\rm max}$ weighting to construct the number density distributions of $g-r$ color and sSFR for the void and non-void galaxy samples. The results are shown in Figure~\ref{fig:ndd}. We find that even after controlling for stellar mass, there are significant differences between void and non-void galaxies. Specifically, void galaxies are consistently bluer and exhibit higher sSFR compared to their non-void counterparts across the mass range. These results suggest that the observed differences in galaxy properties cannot be attributed solely to intrinsic properties linked to their stellar mass, but reflect a genuine environmental effect.

\begin{figure*}
    \centering
    \includegraphics[width=\textwidth]{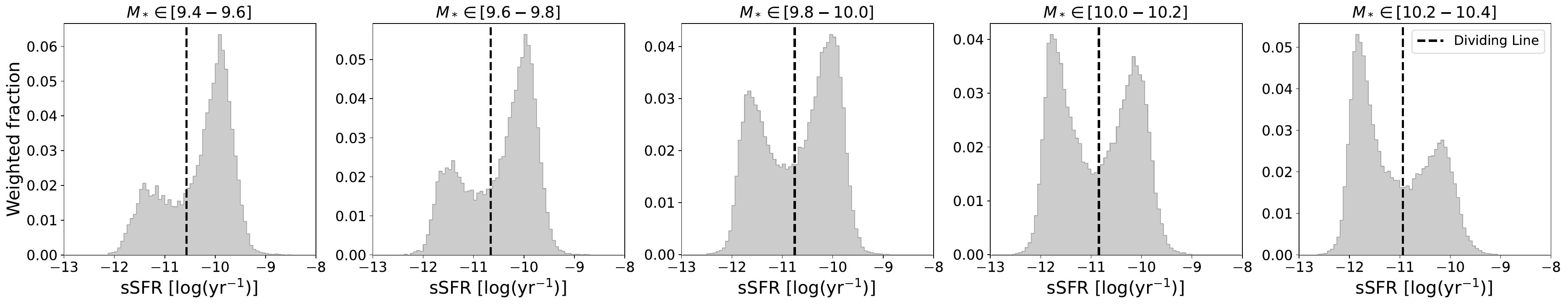}
    \caption{Classification of galaxies into star-forming and quiescent populations across five stellar mass bins based on the normalized number density distribution of sSFR. The gray shaded histogram represents the distribution of all galaxies from our SDSS DR7 catalog. The black vertical dashed line marks the dividing line that separates the star-forming and quiescent populations.}
    \label{fig:ssfrdl}
\end{figure*}

\subsection{Quantifying Environmental Effects with $g-r$ Color and sSFR}\label{sec:qe}

Having demonstrated that the differences between void and non-void galaxies persist after controlling for stellar mass, we present a quantitative assessment of these environmental effects. In this section, we further divide the galaxy sample within each stellar mass bin based on two key properties.

For color, we separate galaxies into blue and red populations according to their $g-r$ distribution. And for star formation activity, we classify the sample into star-forming and quiescent galaxies based on the sSFR. These divisions allow us to compute the blue cloud to red sequence ratio and the star-forming to quiescent ratio for void and non-void galaxies separately. By comparing these ratios, we can directly measure how strongly the void environment affects galaxy properties.

\subsubsection{The Blue Cloud to Red Sequence Ratio}\label{sec:b/r}

To classify galaxies into blue cloud and red sequence populations, we apply the Gaussian bimodal fitting to the all galaxy sample in our catalog within each of the five stellar mass bins from 9.4 to 10.4 in $\log[M_*/\mathrm{M}_\odot]$, which models the $g-r$ color distribution as the sum of two Gaussian components representing the blue and red populations. The derived dividing lines for $g-r$ in these five mass bins, chosen as the crossover between the fitted blue and red distributions, are $0.628 \pm 0.004, 0.644 \pm 0.003, 0.671 \pm 0.005, 0.685 \pm 0.002, \text{ and } 0.694 \pm 0.001$, with $1\sigma$ uncertainties estimated via the jackknife resampling method using 100 sub-samples. The relative errors of these dividing lines are less than 1\%, which do not affect our subsequent analysis. 
The results of the $g-r$ color bimodal distribution are shown in Figure~\ref{fig:grdl}.

Based on this classification, we compute the blue cloud to red sequence ratio, $R_{\rm b/r}$, for the number density distributions of void and non-void galaxies separately within each mass bin (see Figure~\ref{fig:ndd}), where the ratio is defined as the total number density of blue galaxies to that of red galaxies. To estimate the uncertainties, we apply the jackknife resampling method by dividing the full sample into 100 sub-samples. In Figure~\ref{fig:rbr}, we show $R_{\rm b/r}$ for void and non-void galaxies separately, along with the ratio between these two values, defined as $\mathcal{R}_{\rm b/r} \equiv R_{\rm b/r}^{\rm void} / R_{\rm b/r}^{\rm non-void}$. The values of $\mathcal{R}_{\rm b/r}$ we obtain are $2.057\pm0.167$, $1.775\pm0.115$, $1.433\pm0.065$, $1.164\pm0.044$, and $1.419\pm0.056$, respectively. 
We perform a weighted linear fit to these values, which yields a slope of $-0.5956 \pm 0.1472$ per dex. The negative slope indicates an overall decreasing trend with stellar mass, confirming that void environments have a measurable effect on galaxy properties, and that the strength of this effect depends on stellar mass.

\subsubsection{The Star-forming to Quiescent Ratio}\label{sec:s/q}

\begin{figure}
\centering
\includegraphics[width=\columnwidth]{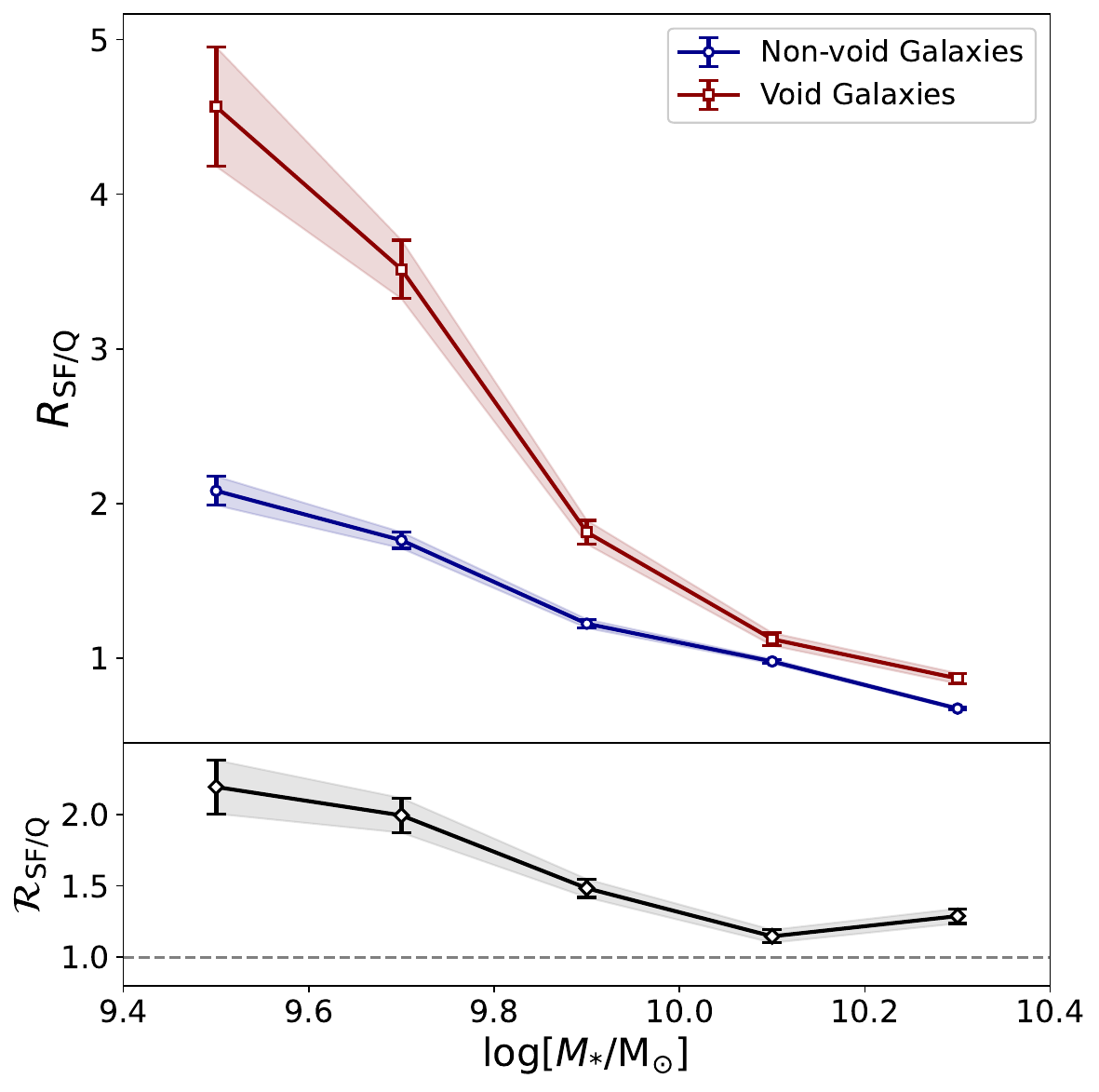}
\caption{The ratio of star-forming to quiescent galaxies \(R_{\rm SF/Q}\) as a function of stellar mass for void galaxies (dark red) and non-void galaxies (dark blue). The bottom panel shows the ratio of these two values, \(\mathcal{R}_{\rm SF/Q}\). The error bars and shaded regions indicate the \(1\sigma\) uncertainties.}
\label{fig:rsfq}
\end{figure}

The sSFR distribution exhibits a bimodal structure separated by a valley between the star-forming and quiescent populations. We fit this valley as a linear function of stellar mass:
\begin{equation}
\log[\mathrm{sSFR} / \mathrm{yr}^{-1}] = -0.46 \times \log[M_* / \mathrm{M}_\odot] - 6.2.
\end{equation}
To separate galaxies into star-forming and quiescent populations, we adopt the resulting dividing line based on this relation.
In Figure~\ref{fig:ssfrdl}, we show the classification of the all galaxy sample of our SDSS DR7 catalog based on this relation, within each of the five stellar mass bins from 9.4 to 10.4 in $\log[M_*/\mathrm{M}_\odot]$.

For each mass bin, we then calculate the star-forming to quiescent ratio, $R_{\rm SF/Q}$, using the number density distribution of void and non-void galaxies in each mass bin individually (see Figure~\ref{fig:ndd}), defined as the ratio of the total number density of star-forming galaxies to that of quiescent galaxies. Uncertainties are derived using the same jackknife resampling method with 100 sub-samples. In Figure~\ref{fig:rsfq}, we show $R_{\rm SF/Q}$ for void and non-void galaxies, along with the ratio between these two quantities, defined as $\mathcal{R}_{\rm SF/Q} \equiv R_{\rm SF/Q}^{\rm void} / R_{\rm SF/Q}^{\rm non-void}$. The measured values of $\mathcal{R}_{\rm SF/Q}$ are $2.192\pm0.188$, $1.993\pm0.121$, $1.483\pm0.064$, $1.147\pm0.045$, and $1.288\pm0.052$, respectively. 
We perform a weighted linear fit to these values, which yields a slope of $-0.9546 \pm 0.1485$ per dex. The negative slope indicates an overall decreasing trend with stellar mass, in agreement with the trend found in the $g-r$ analysis.

The ratios $\mathcal{R}_{\rm b/r}$ and $\mathcal{R}_{\rm SF/Q}$ quantify the strength of the environmental effect by comparing the blue to red and star forming to quiescent ratios of void galaxies to those of non-void galaxies. Their overall decreasing trend with stellar mass demonstrates again that void environments have a significant effect on galaxy evolution, with a stronger impact in lower-mass systems.

\section{Conclusion}
\label{sec:conclusion}

In this work, we focus on the large-scale density environmental effects on galaxy evolution, and we investigated the influence of cosmic void and non-void environments on galaxy properties using the SDSS DR7 galaxy catalog.
We first identified non-spherical voids via Voronoi tessellation and the watershed algorithm, and classified void galaxies based on their local volume relative to the mean value. Using this classification, we compared the number distributions of stellar mass, luminosity, color, and sSFR for void galaxies with those of non-void galaxies. 
Subsequently, to reduce the influence of intrinsic stellar mass effects\footnote{Note that the host halo mass, morphology, and redshift of galaxies could also affect their galaxy properties, such as $g-r$ color and sSFR. If we aim to completely separate the environmental effect from the intrinsic properties, controlling for all these parameters would need to be considered.}, we applied the $V_{\rm max}$ method to construct the number density distributions of $g-r$ and sSFR, and divided the galaxy samples into different stellar mass bins. 
We then further classified galaxies into blue and red populations based on their $g-r$ color, and into star-forming and quiescent populations based on their sSFR. This allowed us to quantify the environmental dependence using the ratio of blue to red galaxies ($R_{\rm b/r}$) and the ratio of star-forming to quiescent galaxies ($R_{\rm SF/Q}$).

Our results show that void galaxies classified from non-spherical voids using our method are on average less massive, fainter, bluer, and have higher sSFR than non-void galaxies. To further quantify this environmental dependence, we examined $R_{\rm b/r}$ and $R_{\rm SF/Q}$ derived from the number density distributions within five stellar mass bins over the $9.4-10.4$ range in $\log[M_*/\mathrm{M}_\odot]$. Both ratios are higher for void galaxies than for non-void galaxies at each stellar mass bin. When we take the ratio of the void to the non-void value for $R_{\rm b/r}$ and $R_{\rm SF/Q}$, we find that both quantities exhibit an overall decreasing trend with stellar mass. These results provide a quantitative measurement of the environmental effect associated with non-spherical voids.
These results suggest that in underdense regions, galaxies experience weaker gravitational interactions and less frequent gas exchange with their surroundings compared to those in denser environments. Galaxies in voids are found to assemble their stars slowly \citep{2023Natur.619..269D}, which may allow them to remain in an earlier, more active evolutionary stage, resulting in bluer colors and higher star formation rates.

Overall, the method of classifying void galaxies based on their local volume in non-spherical voids provides a reliable way to identify galaxies in low-density environments. Furthermore, the method based on the ratios of blue to red galaxies and star-forming to quiescent galaxies allows us to measure the environmental effect in a robust and quantitative manner, directly demonstrating how low-density environments influence galaxy evolution. These findings demonstrate the effectiveness of using non-spherical voids as a powerful probe for investigating the environmental dependence of galaxy properties.

\begin{acknowledgments}

This work is supported by the National Key R\&D Program of China (2023YFA1607800, 2023YFA1607804, 2022YFA1605300), the China Manned Space Project with Nos. CMS-CSST-2021-A02 \& CMS-CSST-2025-A04, “the Fundamental Research Funds for the Central Universities”, 111 project No. B20019, and Shanghai Natural Science Foundation, grant No. 19ZR1466800, and the National Nature Science Foundation of China
(NSFC) grants No. 12273051. This project is also supported in part by Office of Science and Technology, Shanghai Municipal Government (grant Nos. 24DX1400100, ZJ2023-ZD-001). Y.G. acknowledges the support from the CAS Project for Young Scientists in Basic Research (No. YSBR- 92), National Key R\&D Program of China grant Nos. 2022YFF0503404, and science research grants from the China Manned Space Project with grant Nos. CMS-CSST-2025-A02. H.T.M. acknowledges support from the National Natural Science Foundation of China (NSFC, grant No. 12503008). Y.Z.G. acknowledges the support of the National Natural Science Foundation of China under the grant number 12503013. Y.J.P. acknowledges support from the National Natural Science Foundation of China (NSFC) under grant Nos. 12125301 and 12192222, and from the New Cornerstone Science Foundation through the XPLORER PRIZE.This work has made use of the Gravity Supercomputer at the Department of Astronomy, Shanghai Jiao Tong University.

\end{acknowledgments}


\appendix

\section{Normalized number distributions of galaxy properties for void galaxy samples with $f=1,2,3$}\label{sec:appendix1}

\begin{figure*}
    \centering
    \subfigure{\includegraphics[width=\columnwidth]{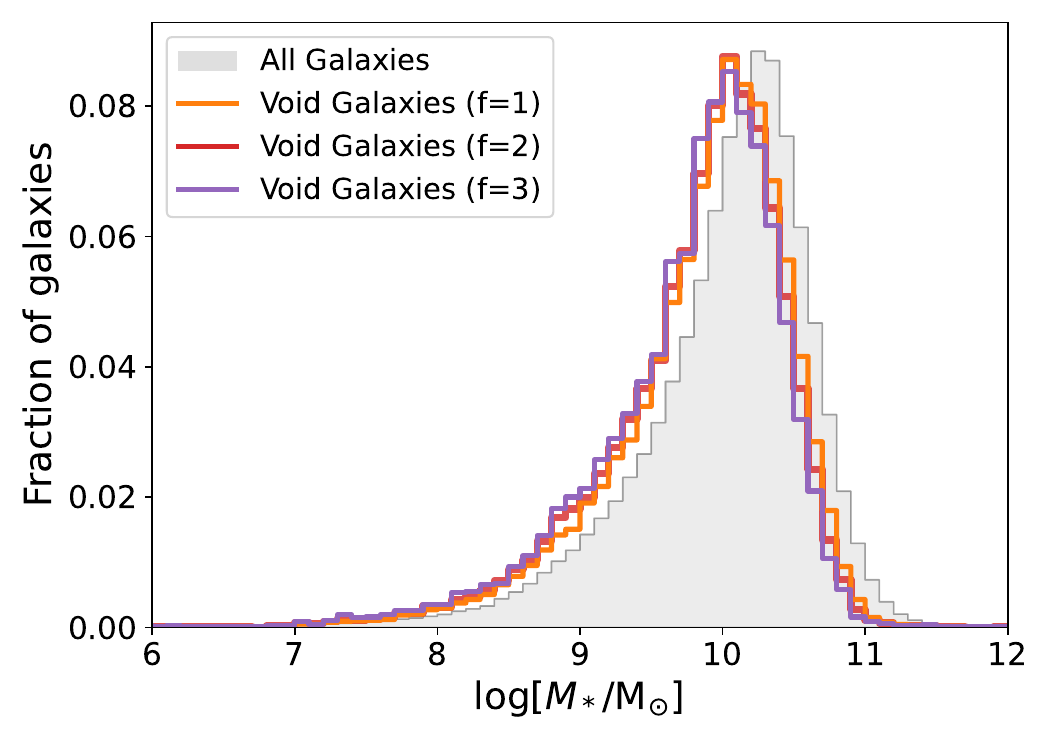}}
    \hspace{2mm}
    \subfigure{\includegraphics[width=\columnwidth]{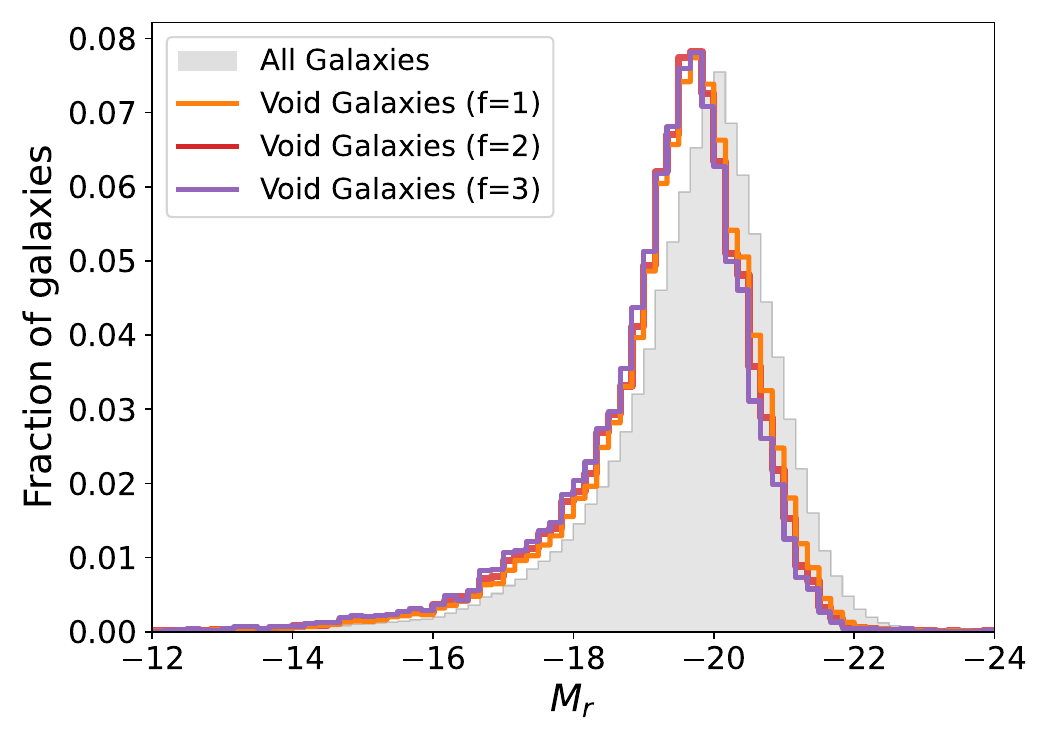}}
    \subfigure{\includegraphics[width=\columnwidth]{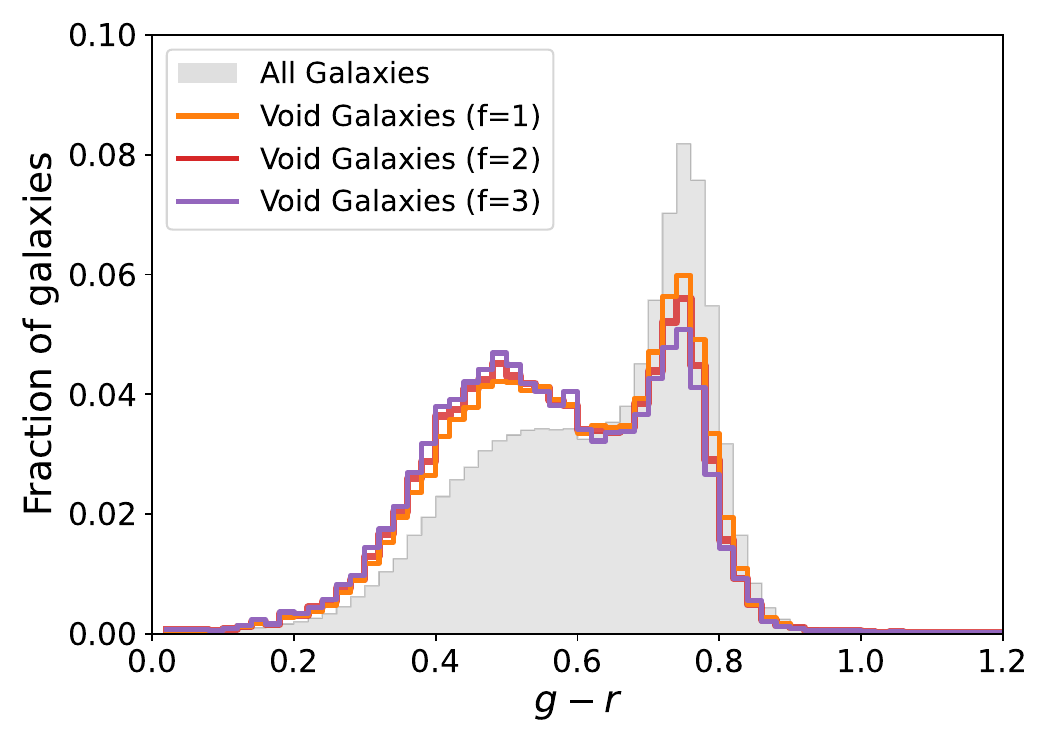}}
    \hspace{2mm}
    \subfigure{\includegraphics[width=\columnwidth]{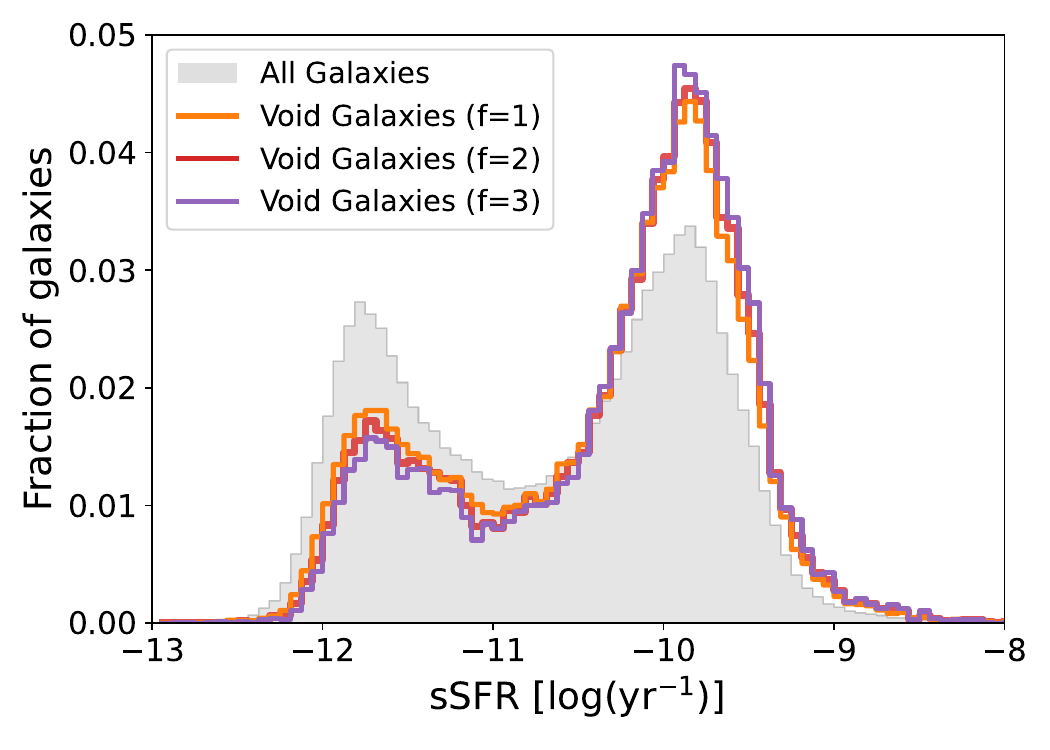}}
    \caption{The normalized number distributions of $M_*$, $M_r$, $g-r$, and sSFR for the void galaxy samples with $f=1$ (orange), $f=2$ (red), and $f=3$ (purple), together with the all galaxy sample (gray shaded). Each distribution is normalized by the total number of galaxies in the corresponding sample.}\label{fig:vnda}
\end{figure*}

As shown in Figure~\ref{fig:vnda}, we present the normalized number distributions of $M_*$, $M_r$, $g-r$, and sSFR for the void galaxy samples with $f=1, 2, 3$, where the normalisation is by the total number of galaxies in each sample. Clear offsets between the void galaxy samples for all three $f$ values and the all galaxy sample are seen in all four panels.

\section{Void Galaxy Identification from Spherical Voids}\label{sec:appendix2}

\begin{figure*}
    \centering
    \subfigure{\includegraphics[width=\columnwidth]{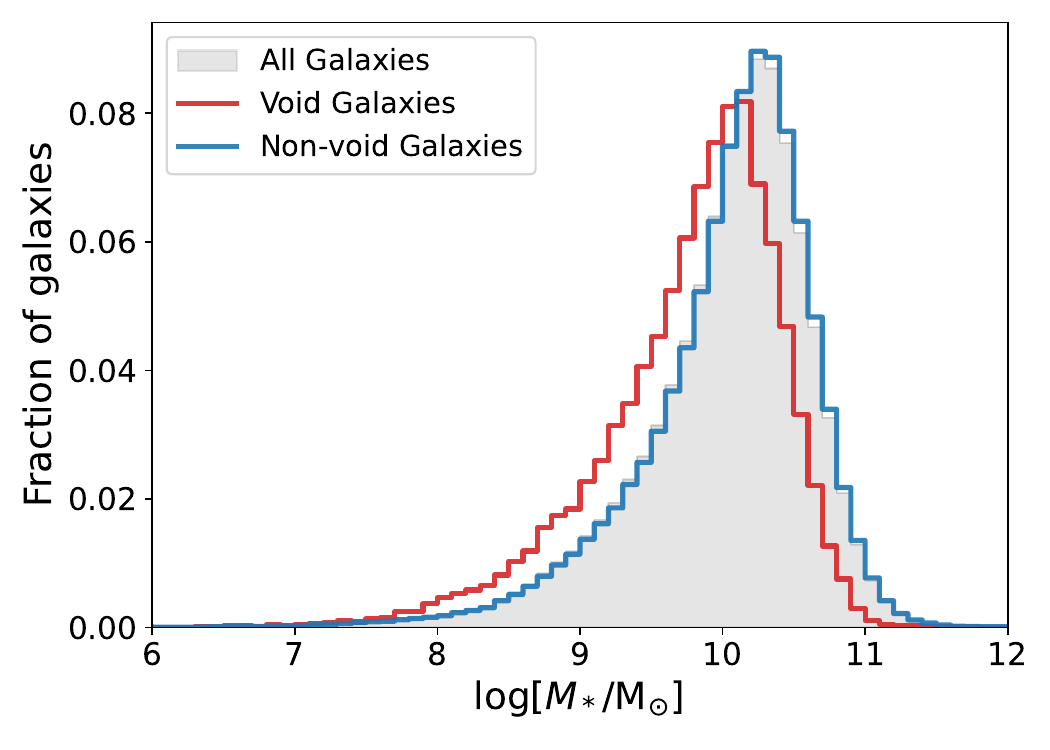}}
    \hspace{2mm}
    \subfigure{\includegraphics[width=\columnwidth]{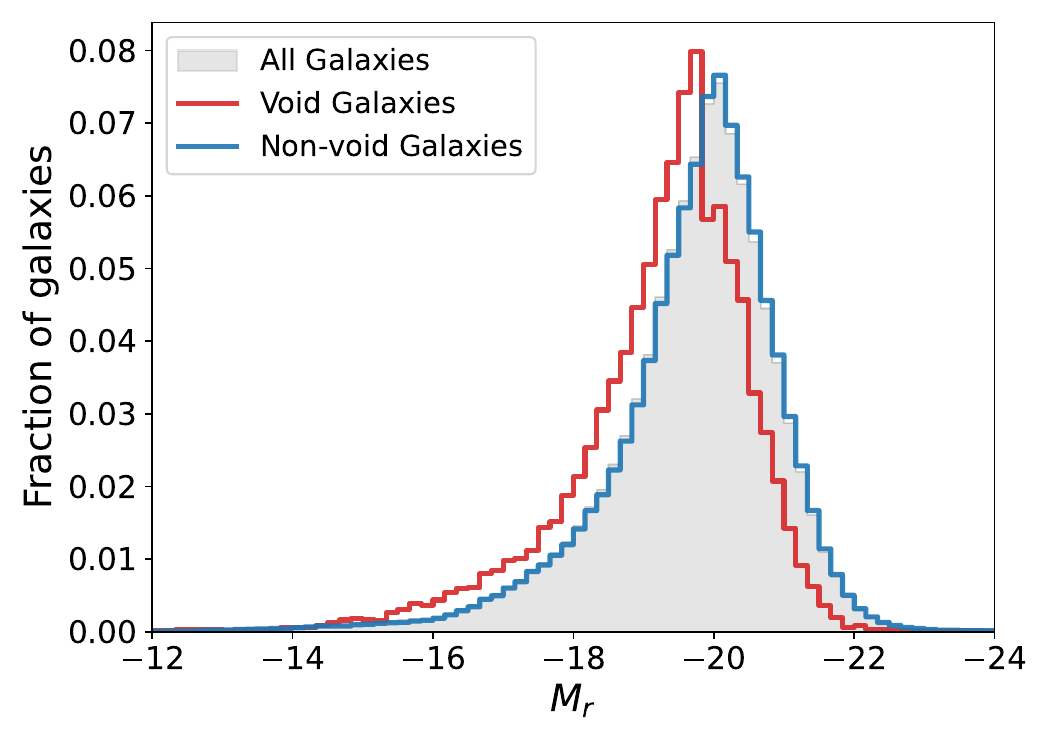}}
    \subfigure{\includegraphics[width=\columnwidth]{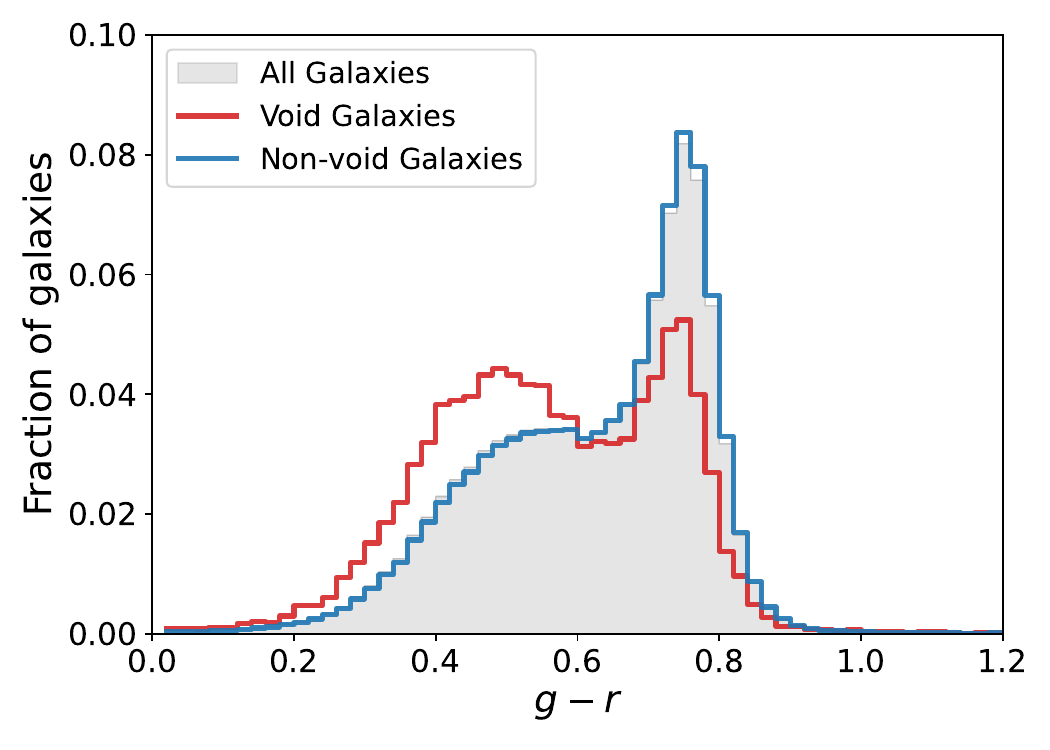}}
    \hspace{2mm}
    \subfigure{\includegraphics[width=\columnwidth]{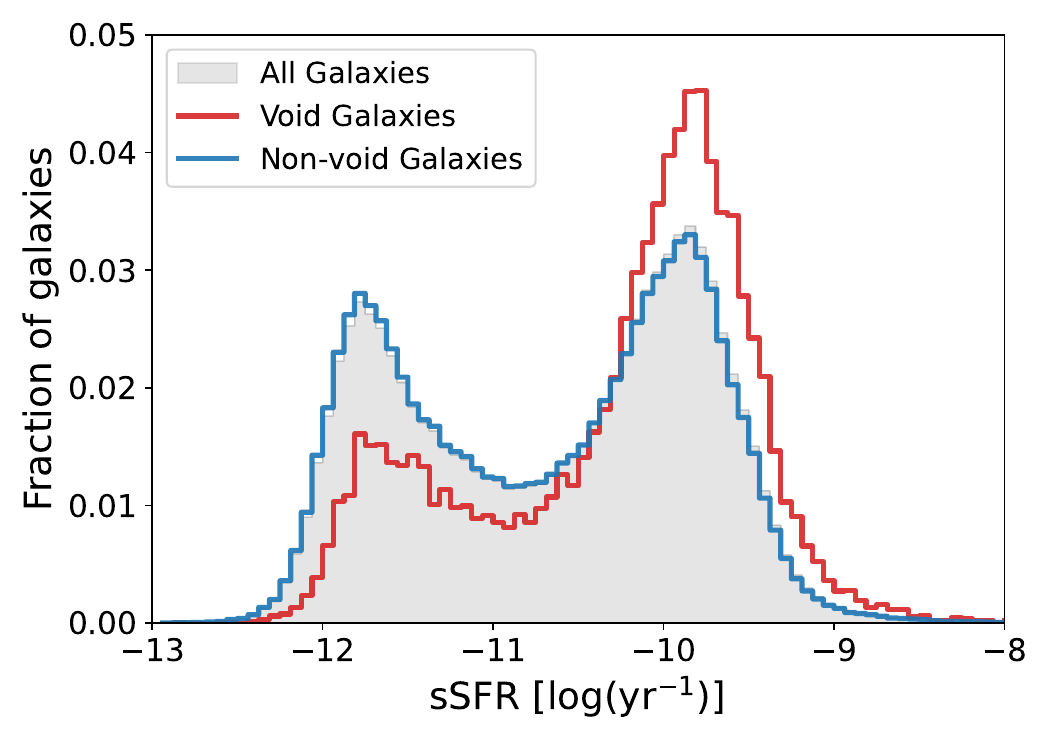}}
    \caption{The normalized number distributions of $M_*$, $M_r$, $g-r$, and sSFR for the void galaxy sample classified from the spherical void catalog (red), non-void sample (blue), and the all galaxy sample (gray shaded). Each distribution is normalized by the total number of galaxies in the corresponding sample.}\label{fig:svnd}
\end{figure*}

We identify spherical voids from a volume-limited subsample of our SDSS DR7 galaxy catalog with $M_r \leq -20$ at $z = 0-0.114$, which contains $\sim 194,000$ galaxies. This volume-limited sample is adopted to be consistent with previous work \citep[e.g.][]{2025ApJ...978....3Z}. For the spherical void identification, we use the \texttt{VoidFinder} algorithm \citep{1997ApJ...491..421E,2002ApJ...566..641H} from the Void Analysis Software Toolkit\footnote{\url{https://github.com/desi-ur/vast}} \citep[\texttt{VAST},][]{vast}. \texttt{VoidFinder} identifies voids by growing and merging empty spheres, and we retain only those with radii larger than $10 \, h^{-1}{\rm Mpc}$. The spherical void catalog we use contains approximately 1150 voids with a mean radius of $12.4 \, h^{-1}{\rm Mpc}$.

A galaxy is classified as a void galaxy if its position falls within one of these spherical voids, and this classification identifies a sample of 23,018 void galaxies from our SDSS DR7 catalog. The normalized number distributions of $M_*$, $M_r$, $g-r$, and sSFR for this sample are shown in Figure~\ref{fig:svnd}, together with the corresponding non-void sample and the all galaxy sample, and we notice that the median values of $M_*$, $M_r$, $g-r$, and sSFR distributions for this void galaxy sample are 9.793, $-19.630$, 0.537, and $-10.213$, respectively.

\bibliography{voidgalaxy}
\bibliographystyle{aasjournal}

\end{document}